\documentstyle[12pt]{article}

\begin{document}

\renewcommand{\theequation}{\thesection.\arabic{equation}}
\renewcommand{\thefootnote}{\fnsymbol{footnote}}

\hsize37truepc\vsize61truepc
\hoffset=-.5truein\voffset=-0.8truein
\setlength{\baselineskip}{17pt plus 1pt minus 1pt}
\setlength{\textheight}{22.5cm}

\def\vin{v^{\rm in}}
\def\vout{v^{\rm out}}
\def\hin{h^{\rm in}}
\def\hout{h^{\rm out}}
\def\iso{\simeq}
\def\End{{\rm End}}
\def\diag{{\rm diag}}
\def\I{{\rm i}}
\def\tr{{\rm tr}}
\def\taux{t^{\rm aux}}
\def\boldsp{\mbox{\boldmath $\omega$}^A}
\def\boldalpha{\mbox{\boldmath $\alpha$}}
\def\boldbeta{\mbox{\boldmath $\beta$}}
\def\bigboldsigma{\mbox{\boldmath $\sigma$}}
\def\boldsigma{\mbox{\small \boldmath $\sigma$}}
\def\smallboldalpha{\mbox{\small \boldmath $\alpha$}}
\def\smallboldbeta{\mbox{\small \boldmath $\beta$}}
\def\boldomega{\mbox{\boldmath $\omega$}}
\def\rlx{\relax\leavevmode}
\def\inbar{\vrule height1.5ex width.4pt depth0pt}
\def\IC{\rlx\hbox{\,$\inbar\kern-.3em{\rm C}$}}
\def\IZ{\rlx\hbox{\,$\inbar\kern-.3em{\rm Z}$}}
\def\smallfrac#1#2{\mbox{\small $\frac{#1}{#2}$}}
\def\lem#1{{\it Lemma #1. }}
\def\prop#1{{\it Proposition #1. }}
\def\E{{\rm e}}

\newtheorem{proposition}{Proposition}
\newtheorem{lemma}{Lemma}
\newtheorem{conjecture}{Conjecture}
\newtheorem{claim}{Claim}

\begin{titlepage}

\noindent January, 1995 \hfill{MRR 014-95}\\
\mbox{ } \hfill{hep-th/9502041}
\vskip 1.6in
\begin{center}
{\Large {\bf Exact solution for the spin-$s$ XXZ quantum chain}}\\[8pt]
{\Large {\bf with non-diagonal twists}}
\end{center}

\normalsize
\vskip .4in

\begin{center}
C. M. Yung  \hspace{3pt}
and \hspace{3pt} M. T. Batchelor
\par \vskip .1in \noindent
{\it Department of Mathematics, School of Mathematical Sciences}\\
{\it Australian National University, Canberra ACT 0200, Australia}
\end{center}
\par \vskip .3in

\begin{center}
{\Large {\bf Abstract}}\\
\end{center}

We study integrable vertex models and quantum spin chains with toroidal
boundary conditions. An interesting class of such boundaries is associated
with non-diagonal twist matrices. For such models there are no trivial
reference states upon which a Bethe ansatz calculation can be constructed,
in contrast to the well-known case of periodic boundary conditions. In
this paper we show how the transfer matrix eigenvalue expression
for the spin-$s$ XXZ chain
twisted by the charge-conjugation matrix can in fact be obtained.
The technique used is the generalization to spin-$s$ of the functional
relation method based on ``pair-propagation through a vertex''. The Bethe
ansatz-type equations obtained reduce, in the case of lattice size $N=1$, to
those recently found for the Hofstadter problem of Bloch electrons on a
square lattice in a magnetic field.


\vspace{1cm}
\noindent
{\bf Running title}: Spin-$s$ XXZ chain with twists

\end{titlepage}

\section{Introduction}

Vertex models and related quantum spin chains are most often studied with
periodic boundary conditions \cite{Baxter82,Faddeev79,Kulish82}. If the
Boltzmann weights underlying the models are obtained from solutions of the
Yang-Baxter equation, or $R$-matrices, then commutativity of the associated
transfer matrices follows and leads to integrability.
A large class of $R$-matrices and therefore
integrable vertex models with periodic boundaries can be constructed via the
machinery of quantum groups \cite{Jimbo86,Bazhanov87}. Several techniques --
coordinate Bethe ansatz, algebraic Bethe ansatz, analytic Bethe ansatz, etc.\
-- have been developed for diagonalizing the corresponding transfer matrices.

More general toroidal boundary conditions than periodic are in fact
integrable; the condition being that the ``twist matrix'' specifying the
boundary condition is a ``symmetry'' of the $R$-matrix \cite{deVega84}.
Especially interesting are the cases where the twist matrix is non-diagonal.
In spite of the existence of commuting transfer matrices for such cases,
their diagonalization by the above mentioned Bethe ansatz
techniques seems to fail. This is due to
the non-existence in such cases of a trivial reference state.

In this paper we show that the spin-$s$ XXZ chain twisted by the (non-diagonal)
charge-conjugation matrix can in fact be diagonalized. The technique used is a
functional relation method \cite{Baxter82} (also known as the ``T-Q''
method or the ``method of commuting transfer matrices'') based on the
``pair-propagation
through a vertex'' property of the relevant $R$-matrices. This generalizes
to spin-$s$ a recent result for the spin-$\smallfrac{1}{2}$ chain
(or six-vertex model) \cite{Batchelor95}.

One way \cite{Kulish82,Kirillov87} to study the spin-$s$ XXZ chain
with periodic boundaries is to first express the transfer matrix
$t^{(2s)}_{2s}(u)$ in terms of a much simpler auxiliary transfer matrix
$t^{(2s)}_1(u)$ using the fusion procedure \cite{Kulish82}. This auxiliary
transfer matrix is then diagonalized with the algebraic Bethe ansatz to give
the eigenvalue expression and associated Bethe ansatz equations for the
original spin-$s$ chain. We show that in the presence of the twist, there is
an analogous procedure. The corresponding auxiliary transfer matrix is then
diagonalized with the abovementioned functional relation method.
As a consquence we obtain the transfer matrix eigenspectrum for the
spin-$s$ XXZ chain with the charge-conjugation twist.

The paper is organized as follows: Section 2 is introductory in nature. We
describe the general theory of integrable toroidal boundary conditions here
and make a survey of available results for the case of the
spin-$\smallfrac{1}{2}$ XXZ Heisenberg chain. In Section 3 we study the
integrable spin-$s$ generalization of the XXZ chain twisted by the charge
conjugation matrix. In particular we relate the associated transfer
matrix $t^{(2s)}_{2s}(u)$ to the auxiliary
transfer matrix $t^{(2s)}_1(u)$, which is defined in terms of the $R$-matrix
$R^{(2s,1)}(u)$. Section 4 is the body of the paper. In
Section 4.1 we show that the property of ``pair-propagation through a
vertex'' for the $R$-matrix $R^{(1,1)}(u)$ and related spin-$\smallfrac{1}{2}$
XXZ chain \cite{Baxter82} generalizes to arbitrary spin-$s$. This allows the
$T-Q$ relations for this auxiliary
transfer matrix to be obtained in Section 4.2. For this purpose we required
a conjectured property for two families of matrices $Q_L(u)$ and $Q_R(u)$,
which is partially proved in Appendix A. In Section 4.3 we obtain the
eigenvalue expression and Bethe ansatz-type relations for the auxiliary
transfer matrix, leading to the corresponding expressions for the spin-$s$
XXZ chain. In Section 5 we discuss the contents of this paper in a general
context.

After this work was completed, we became aware of an interesting connection
between integrable non-diagonal twists for the auxiliary transfer matrix
$t^{(2s)}_1(u)$ in the case of lattice size $N=1$ and the so-called
Hofstadter problem of Bloch electrons on a square lattice in a magnetic
field \cite{Wiegmann94}. In Appendix B we make contact with the latter
work; in particular, recovering their Bethe ansatz results derived using
a completely different method.

\section{Integrable models with toroidal boundary conditions}
\setcounter{equation}{0}

Let us recall a few fundamental facts concerning the quantum inverse
scattering method and the construction of integrable models
\cite{Faddeev79,Kulish82}. The principal ingredient is the Yang-Baxter
equation
\begin{equation}
R_{12}(u) R_{13}(u+v) R_{23}(v) = R_{23}(v) R_{13}(u+v) R_{12}(u)
\label{eqn:ybe}
\end{equation}
which is an equation in the space $V_1 \otimes V_2 \otimes V_3$, with
$V_i \equiv V \iso \IC^n$, $R(u) \in \End(V\otimes V)$ and $R_{12}(u)=
R(u)\otimes 1$, $R_{23}(u)=1\otimes R(u)$, etc. Given an $R$-matrix, which
by definition is a solution of (\ref{eqn:ybe}) one constructs a monodromy
matrix
\begin{equation}
\stackrel{a}{T}(u) = R_{a1}(u) R_{a2}(u) \cdots R_{aN}(u),
\label{eqn:monod}
\end{equation}
which is to be thought of as an operator-valued matrix in the
auxiliary space $V_a$. A consequence of (\ref{eqn:ybe}) is the
intertwining relation
\begin{equation}
R_{12}(u-v) \stackrel{1}{T}(u) \stackrel{2}{T}(v) =
\stackrel{2}{T}(v) \stackrel{1}{T}(u) R_{12}(u-v)
\label{eqn:int}
\end{equation}
for the monodromy matrices. The indices 1 and 2 here label
different auxiliary spaces and are not to be confused with those appearing
in the definition (\ref{eqn:monod}) which label different quantum spaces.
The important object
\begin{equation}
t_P(u) = \tr_a \; \stackrel{a}{T}(u),
\label{eqn:tmp}
\end{equation}
has the interpretation of a row-to-row transfer matrix for an $n$-state
vertex model with periodic boundary conditions and forms a commuting
family.

We consider an extension of (\ref{eqn:tmp}) to more general toroidal
boundaries; namely by studying the transfer matrix
\begin{equation}
t(u) = \tr_a \; \stackrel{a}{T}(u) \stackrel{a}{F}
\label{eqn:tma}
\end{equation}
where $F$ is a square matrix of dimension ${\rm dim} V_a$. The matrix $F$
is called a ``twist matrix'' and the boundary conditions twisted. The
following simple result shows that for certain twist matrices $F$ commutativity
of transfer matrices is retained:
\begin{proposition}\cite{deVega84}
Let the twist matrix $F$ be a ``symmetry'' of the $R$-matrix, i.e\ it obeys
$[R(u),F \otimes F]=0$.  Then the transfer matrix (\ref{eqn:tma}) satisfies
$[t(u),t(v)]=0$ and we say $F$ is integrable.
\end{proposition}
\noindent
{\bf Proof }
{}From the definition, we have $t(u)\;t(v)=\tr_{12} \stackrel{1}{F}
\stackrel{1}{T}(u) \stackrel{2}{T}(v) \stackrel{2}{F}$. Using the interwining
relation (\ref{eqn:int}) to change the order of $\stackrel{1}{T}(u)
\stackrel{2}{T}(v)$ we obtain
\begin{eqnarray*}
 \tr_{12} \stackrel{1}{F}
\stackrel{1}{T}(u) \stackrel{2}{T}(v) \stackrel{2}{F} & = & \tr_{12}
   \stackrel{1}{F} R_{12}^{-1}(u-v) \stackrel{2}{T}(v) \stackrel{1}{T}(u)
   R_{12}(u-v) \stackrel{2}{F} \\
 & = & \tr_{12} R_{12}^{-1}(u-v) \stackrel{2}{T}(v) \stackrel{1}{T}(u)
        R_{12}(u-v) F \otimes F \\
({\rm symmetry}) & = &  \tr_{12} R_{12}^{-1}(u-v) \stackrel{2}{T}(v)
      \stackrel{1}{T}(u) F \otimes F  R_{12}(u-v) \\
& = & \tr_{12} \stackrel{2}{T}(v) \stackrel{1}{T}(u) \stackrel{1}{F}
   \stackrel{2}{F} \;\; =  \;\; t(v)\;t(u)
\end{eqnarray*}

A quantum spin chain can be associated with the transfer matrix
(\ref{eqn:tma}) in the usual way. For simplicity we assume the $R$-matrices
are regular, with $R(0)=c P$, where $P$ is the exchange operator
with matrix elements $P_{a\;b}^{c\;d}=\delta_a^d \delta_b^c$. Given
the multi-indices $(\alpha_1,\ldots,\alpha_N)$ and $(\beta_1,\ldots,\beta_N)$
we have from (\ref{eqn:tma})
\begin{equation}
t(0)_{\alpha_1\ldots\alpha_N}^{\beta_1\ldots\beta_N} =
  \delta_{\alpha_1}^{\beta_2}
  \cdots \delta_{\alpha_{N-1}}^{\beta_{N}}  F_{\alpha_N}^{\beta_1} c^N.
\end{equation}
Similarly, assuming $F$ is invertible,  we have
\begin{eqnarray}
\left(t(0)^{-1} t'(0)\right)_{\alpha_1\ldots\alpha_N}^{\beta_1\ldots\beta_N}
& = & c^{-1} \left(
  R'(0)^{\beta_2\;\beta_1}_{\alpha_1\;\alpha_2}
  \delta_{\alpha_3}^{\beta_3}\cdots \delta_{\alpha_{N}}^{\beta_{N}}+
\cdots \right.\nonumber\\
& & \cdots + \delta_{\alpha_1}^{\beta_1}\cdots
   \delta_{\alpha_{N-2}}^{\beta_{N-2}}
   R'(0)^{\beta_{N}\;\beta_{N-1}}_{\alpha_{N-1}\;\alpha_{N}}+\nonumber\\
& &  \left.\delta_{\alpha_2}^{\beta_2}\cdots
\delta_{\alpha_{N-1}}^{\beta_{N-1}}
  (F^{-1})_{\alpha_1}^{\epsilon}
  R'(0)^{\eta\;\beta_N}_{\alpha_N\;\epsilon} F_{\eta}^{\beta_1}\right)
\end{eqnarray}
{}From this we conclude that the quantum spin chain has Hamiltonian
\begin{equation}
H\equiv t(0)^{-1} t'(0)= \frac{1}{c}\left(\sum_{j=1}^{N-1} h_{j,j+1} +
h^{\rm F}_{N,1}\right),
\label{eqn:ham}
\end{equation}
with
\begin{equation}
h_{i,j}=R_{ij}'(0) {\cal P}_{ij}, \hspace{20pt}
h^{\rm F}_{N,1}= \stackrel{1}{F^{-1}} h_{N,1} \stackrel{N}{F},
\end{equation}
where $\stackrel{p}{F}$ acts non-trivially only in the $p$-th slot of $V_1
\otimes \cdots \otimes V_N$. When the twist matrix $F$ is the identity,
the spin chain with periodic b.c.'s is recovered.

\vspace{5pt}
\noindent {\bf Example: Spin-$\smallfrac{1}{2}$ XXZ chain}
\vspace{5pt}

\noindent
The six-vertex model $R$-matrix is given by
\begin{displaymath}
R(u) = \left(\begin{array}{cccc}
 \sinh(u+\lambda) & & & \\
  & \sinh u &  \sinh(\lambda) &\\
  & \sinh(\lambda) & \sinh u & \\
  & & & \sinh(u+\lambda) \end{array}\right)
\end{displaymath}
and admits the following two types of integrable twist matrices:
\begin{displaymath}
({\rm i}) \;\; F= \left( \begin{array}{cc}
     \alpha & 0\\
     0 & \beta\end{array} \right),\hspace{30pt}
({\rm ii})\;\; F= \left( \begin{array}{cc}
     0 & \alpha\\
     \beta & 0\end{array} \right),
\end{displaymath}
where $\alpha,\beta$ are completely arbitrary. When $\alpha,\beta$ are
non-zero,\footnote{When one of $\alpha,\beta$ is allowed to be zero, the
vertex model has {\em fixed} boundaries; it does not seem possible to obtain
a local Hamiltonian in this case. Furthermore, the eigenvalue expression
for such type (ii) twists is unknown.}
the corresponding spin-$\smallfrac{1}{2}$ XXZ Hamiltonian is given by
\begin{equation}
H= \frac{1}{\sinh(\lambda)} \sum_{j=1}^{N} \sigma_j^+ \sigma_{j+1}^- +
   \sigma_j^- \sigma_{j+1}^+ + \frac{\cosh(\lambda)}{2}
   \sigma_j^z \sigma_{j+1}^z,
\end{equation}
with boundary conditions
\begin{eqnarray}
({\rm i}) & & \sigma_{N+1}^+ = \frac{\beta}{\alpha} \sigma_1^+, \hspace{5pt}
  \sigma_{N+1}^-  = \frac{\alpha}{\beta}\sigma_1^-, \hspace{5pt}
  \sigma_{N+1}^z  = \sigma_1^z, \nonumber\\
({\rm ii}) & & \sigma_{N+1}^+ = \frac{\beta}{\alpha}\sigma_1^-, \hspace{5pt}
    \sigma_{N+1}^-  =\frac{\alpha}{\beta} \sigma_1^+, \hspace{5pt}
    \sigma_{N+1}^z  = -\sigma_1^z.
\label{eqn:bchalf}
\end{eqnarray}

For boundary conditions of type (i), the total spin operator $\sum_{j=1}^N
\sigma^z_j$ commutes with the Hamiltonian; the trivial reference state is
available and the corresponding transfer $t^{(i)}(u)$
matrix can be diagonalized with the algebraic Bethe ansatz \cite{deVega84},
with very little difference from the periodic case. The transfer matrix
eigenvalue expression is given by
\begin{eqnarray}
\Lambda^{({\rm i})}(u) & = & \alpha \sinh(u+\lambda)^N \prod_{j=1}^n
\frac{\sinh(u-u_j-\lambda)} {\sinh(u-u_j)} \nonumber\\
& + & \beta \sinh(u)^N \prod_{j=1}^n \frac{\sinh(u-u_j+\lambda)}{\sinh(u-u_j)}
\end{eqnarray}
in the sector with total spin $N-2n$,
with the Bethe ansatz equations for $u_k$ determined from analyticity of
$\Lambda^{({\rm i})}(u)$.

On the other hand, boundary conditions of type (ii) leave
only a $\IZ_2$-invariance in the eigenspectrum. For the case
$\alpha=\beta=1$\footnote{We refer to this as the charge-conjugation twist}
(which can be considered as anti-periodic boundary
conditions on the vertex model),
the problem was recently solved \cite{Batchelor95} using the
functional equation method of Baxter \cite{Baxter82}. The transfer matrix
$\tilde{t}(u)$ has eigenvalue expression
\begin{eqnarray}
\tilde{\Lambda}(u) & = &
  \sinh(u+\lambda)^N \prod_{j=1}^N \frac{\sinh[\smallfrac{1}{2}
  (u-u_j-\lambda)]}{\sinh[\smallfrac{1}{2}(u-u_j]} \nonumber\\
& - &
  \sinh(u)^N \prod_{j=1}^N \frac{\sinh[\smallfrac{1}{2}(u-u_j+\lambda)]}
  {\sinh[\smallfrac{1}{2}(u-u_j]}.
\end{eqnarray}
General boundary conditions of type (ii) are in fact
very closely related. This is due to the relation
\begin{equation}
  \det\left( t^{({\rm ii})}(u) - \sqrt{\alpha\beta}\lambda\right)
  \propto \det\left( \tilde{t}(u) -  \lambda\right)
\end{equation}
between the characteristic polynomials for their corresponding transfer
matrices, leading to the result
\begin{equation}
\Lambda^{({\rm ii})}(u) = \sqrt{\alpha\beta} \tilde{\Lambda}(u).
\end{equation}

Type (i) boundary conditions with $\alpha\beta=1$ were studied in Ref.\
\cite{Alcaraz}: At criticality the model gives rise to a conformal field
theory with central charge $c<1$, with the twist parameter determining the
value of $c$.
All the above toroidal boundary conditions have in fact been studied in
Ref.\ \cite{Alcaraz88} (by direct numerical
diagonalization in the case of type (ii)
twists), with conjectures given for their complete operator content.
In their language, the type (ii) boundary conditions share the same physical
properties because they are in the same conjugacy class of $O(2)$, being the
global symmetry of the infinite chain.

\section{Spin-$s$ XXZ quantum chains}
\setcounter{equation}{0}

The integrable spin-$s$ generalization of the XXZ spin chain and related
$R$-matrices have been studied by many authors; amongst them \cite
{Zamolodchikov80,Kulish81,Kulish82,Kulish83,Sogo83,Sogo84,Jimbo85,Babujian85,Kirillov87}
{}.
In particular, the spin-1 Hamiltonian was first constructed in Ref.\
\cite{Zamolodchikov80}.
Bethe ansatz equations for the spin-$s$ models with periodic boundary
conditions
were obtained and studied in Refs.\ \cite{Sogo84,Babujian85,Kirillov87}.
For integrable
boundary conditions associated with diagonal twists, Bethe ansatz equations
were obtained and studied numerically
in Ref.\ \cite{Alcaraz89} and analytically for spin-1 in Ref.\
\cite{Klumper91}.
General toroidal boundary conditions for the spin-1
case was also investigated by direct numerical diagonalization in Ref.\
\cite{Baranowski90}. In this paper, we will mainly follow the approach of Ref.\
\cite{Kirillov87} to construct the integrable spin-$s$ chain.

The transfer matrix for the spin-$s$ XXZ chain is associated with the
$R$-matrix
$R^{(2s,2s)}(u)$ which acts in a tensor product of two spin-$s$ representations
of $U_q(su(2))$. We will study it via its construction by the fusion procedure.
For this purpose we will require the $R$-matrix $R^{(l,m)}(u)$
acting in the tensor product of the spin-$l/2$ and
spin-$m/2$ representations of $U_q(su(2))$ for arbitrary $l,m$. These can all
be
constructed iteratively from $R^{(l,1)}(u)$, whose explicit form
is given by \cite{Kulish83}
\begin{equation}
R^{(l,1)}(u) = \sinh\left(u+\smallfrac{\lambda}{2}
  [l+ S^z \otimes \sigma^z]\right) + \sinh(\lambda) \left[ S^+ \otimes
   \sigma^- + S^- \otimes \sigma^+ \right].
\label{eqn:rl1}
\end{equation}
Here $S^z$ and $S^{\pm}$ are generators for $U_q(su(2))$:
\begin{eqnarray}
[ S^z, S^{\pm}] & = & \pm 2 S^{\pm}, \nonumber\\[0.02in]
[ S^+, S^- ] & = & \frac{\sinh(\lambda S^z)}{\sinh \lambda}.
\end{eqnarray}
In the spin-$l/2$ representation these operators take the forms
$S^z = \diag(l,l-2,\ldots, -l)$ and
\begin{displaymath}
S^+ = \left(S^-\right)^t =
  \frac{1}{\sinh \lambda} \left( \begin{array}{lllll}
  0 & f(1) & & & \\
  & \ddots & \ddots & &\\
  & & \ddots & \ddots &\\
  & & & \ddots & f(l)\\
  & & & & 0 \end{array}\right),
\end{displaymath}
where $f(j)\equiv \sqrt{ \sinh(j \lambda) \sinh [\lambda (l+1-j)]}$. The
operators $\sigma^z$ and $\sigma^{\pm}$ are in the spin-$1/2$ representation.

The $R$-matrix $R^{(l,1)}(u)$ degenerates at $u=\lambda$ and $u=-l \lambda$:
\begin{equation}
R^{(l,1)}(\lambda) = B^{(l)} {\cal P}^{(l+1)}, \hspace{20pt}
R^{(l,1)}(-l\lambda) = C^{(l)} {\cal P}^{(l-1)},
\end{equation}
where $B^{(l)}$ and $C^{(l)}$ are matrices and
${\cal P}^{(l\pm 1)}$ are projection operators onto subspaces of
dimension $l+2$ and $l$, respectively. Using this property and the fusion
procedure \cite{Kulish82} it is possible to establish the key relations
\cite{Kirillov87}

$D_{12}^{(l)} R_{13}^{(l,m)}(u+\lambda) R_{23}^{(1,m)}(u) D_{12}^{(l) \;-1}
= $
\begin{equation}
 \begin{array}{ll}
 \left( \begin{array} {l} R_{(12) 3}^{(l+1,m)}(u) \hspace{100pt} 0\\
   {*} \hspace{20pt} \sinh u \sinh[u+ (m+1)\lambda]
R_{<12> 3}^{(l-1,m)}(u+2\lambda) \end{array} \right) &{\rm if}\;\; l<m,\\[20pt]
\left( \begin{array} {l}\sinh(u+m\lambda) R_{(12) 3}^{(l+1,m)}(u)
     \hspace{60pt} 0\\
   {*} \hspace{20pt} \sinh u \sinh[u+ (m+1)\lambda]
R_{<12> 3}^{(l-1,m)}(u+2\lambda) \end{array} \right) &{\rm if}\;\; l=m,\\[20pt]
 \left( \begin{array} {c}\sinh(u+m\lambda) R_{(12) 3}^{(l+1,m)}(u)
    \hspace{20pt} 0 \\
   {*} \hspace{25pt} \sinh u R_{<12> 3}^{(l-1,m)}(u+2\lambda) \end{array}
  \right) & {\rm if} \;\; l>m.
\end{array}
\label{eqn:DRD}
\end{equation}
Expressions for the matrices $B^{(l)}$, $C^{(l)}$ and $D^{(l)}$ are given
in \cite{Kirillov87}. The spaces $V_{(12)}$ and $V_{<12>}$ in equation
(\ref{eqn:DRD}) are the symmetrized and anti-symmetrized components
respectively of $V_1 \otimes V_2$. The relations (\ref{eqn:rl1}) and
(\ref{eqn:DRD}) allow the $R$-matrix $R^{(l,l)}(u)$, through which we obtain
the spin-$l/2$ XXZ chain, to be obtained recursively. A non-recursive
definition
is also available \cite{Kirillov87}.

Thus far, all our considerations have been purely local in character; in
particular, they are independent of boundary considerations. We will now
specialize to the boundary conditions specified by the non-diagonal twist
matrix (the charge-conjugation twist)
$F^{(l)}=\left(F^{(l)}\right)^{-1}$, which is the $(l+1) \times (l+1)$
matrix with $1$ along the anti-diagonal and $0$ elsewhere. It has the
properties
\begin{equation}
F^{(l)} S^{\pm} F^{(l)} = S^{\mp}, \hspace{20pt}
F^{(l)} S^{z} F^{(l)} = - S^{z}.
\label{eqn:fss}
\end{equation}
Using this it is easy to see from (\ref{eqn:rl1}) that $[R^{(l,1)}(u),
F^{(l)} \otimes F^{(1)}]=0$. The recursion relations (\ref{eqn:DRD}) then
imply that $F^{(l)}$ is a symmetry (in the sense of Proposition 1) of
$R^{(l,l)}(u)$.

Define the monodromy matrices
\begin{equation}
T^{(l)}(u) = R^{(l,2s)}_{a1}(u)\cdots R^{(l,2s)}_{aN}(u)
\end{equation}
with a spin-$\smallfrac{l}{2}$ auxiliary space and a quantum space which is
an $N$-fold tensor product of spin-$s$ representations,
and transfer matrices
\begin{equation}
t^{(2s)}_l(u) = \tr_a T^{(l)}(u) \stackrel{a}{F^{(l)}}.
\label{eqn:tml}
\end{equation}
By Proposition 1 (or rather, the obvious generalization to the case of
non-isomorphic quantum spaces), these form a commuting family: $[t^{(2s)}_l(u),
t^{(2s)}_n(v)]=0$ for all $l,n$.

With respect to the same decomposition as in (\ref{eqn:DRD}) we also have
\begin{equation}
D_{12}^{(l)} \stackrel{1}{F^{(l)}} \stackrel{2}{F^{(1)}} D_{12}^{(l) \; -1}
= \left( \begin{array} {cc}  F^{(l+1)} & 0\\
  0 & - F^{(l-1)} \end{array} \right).
\label{eqn:DFD}
\end{equation}
Consider now the product $t^{(2s)}_l(u+\lambda)t^{(2s)}_1(u)$:
{}From the definition (\ref{eqn:tml}) and commuting the $R$-matrices
appropriately this becomes
\begin{displaymath}
\tr_{ab} \stackrel{a}{F^{(l)}}
\stackrel{b}{F^{(1)}}R^{(l,2s)}_{a1}(u+\lambda)R^{(1,2s)}_{b1}(u)\cdots
R^{(l,2s)}_{aN}(u+\lambda)R^{(1,2s)}_{bN}(u),
\end{displaymath}
which can be written in the form
\begin{displaymath}
\begin{array}{c}
\tr_{ab}\left(D_{ab}^{(l)}\stackrel{a}{F^{(l)}}\stackrel{b}{F^{(1)}}
{D_{ab}^{(l)}}^{-1}\right)\left(D_{ab}^{(l)}R^{(l,2s)}_{a1}(u+\lambda)
R^{(1,2s)}_{b1}(u){D_{ab}^{(l)}}^{-1}\right)\\
\cdots\left(D_{ab}^{(l)}
R^{(l,2s)}_{aN}(u+\lambda)R^{(1,2s)}_{bN}(u){D_{ab}^{(l)}}^{-1}\right).
\end{array}
\end{displaymath}
By application of eqs (\ref{eqn:DRD}) and (\ref{eqn:DFD}) we
arrive at the following result:
\begin{proposition}
The transfer matrices $t^{(2s)}_l(u)$ obey the recursion relations

$t^{(2s)}_l(u+\lambda) t^{(2s)}_1(u) =$
\begin{equation}
 \left\{\begin{array}{ll}
 t_{l+1}^{(2s)}(u) - \sinh u^N \sinh[u+\lambda(2s+1)]^N
  t_{l-1}^{(2s)}(u+2\lambda) & l<2s \\
 \sinh(u+2s\lambda)^N t_{l+1}^{(2s)}(u) - \sinh u^N \sinh[u+\lambda(2s+1)]^N
  t_{l-1}^{(2s)}(u+2\lambda) & l=2s \\
 \sinh(u+2s\lambda)^N t_{l+1}^{(2s)}(u) - \sinh u^N t_{l-1}^{(2s)}(u+2\lambda)
 & l>2s \end{array} \right.
\label{eqn:rec}
\end{equation}
\end{proposition}

Define the shift operator $z$ by $z^{-1} f(u) z = f(u+\lambda)$ for all
functions $f(u)$ and the function $d^{(m)}(u)$ by $d^{(m)}(u)=
\sinh u^N \sinh[u+\lambda(m+1)]^N$. Then the generating function  for the
transfer matrices $t^{(2s)}_l(u)$ is given by
\begin{eqnarray}
\lefteqn{\left[ 1 - z t_1^{(2s)}(u) - z^2 d^{(2s)}(u) \right]^{-1}=}
\hspace{40pt}\nonumber\\
& & \sum_{k=0}^{2s} z^k t_k^{(2s)}(u) + \sum_{l=2s+1}^{\infty} z^k
\prod_{l=2s}^{k-1} \sinh(u+\lambda l)^N t_k^{(2s)}(u).
\label{eqn:gf}
\end{eqnarray}
This can be proven by multiplying both sides by
$\left(1-z t_1^{(2s)}(u) - z^2 d^{(2s)}(u) \right)$ (on the right) and using
the
recursion relations (\ref{eqn:rec}) to show that all coefficients of
$z^k$ for $k \geq 1$ vanish. The corresponding generating function
\cite{Kirillov87} for periodic b.c.'s differ from (\ref{eqn:gf}) in
only the sign of the coefficient of the $z^2$ term on the lhs.

The transfer matrix $t^{(2s)}_{2s}(u)$, out of which the Hamiltonian for
the spin-$s$ XXZ chain is
constructed, can be obtained in terms of the transfer matrix
$t^{(2s)}_1(u)$ either using the recursion relations (\ref{eqn:rec}) or the
generating function (\ref{eqn:gf}). It is this transfer matrix
which is manageable; for instance, because the auxiliary space for
$t^{(2s)}(u)$ is two-dimensional, the functional relation method of Baxter
\cite{Baxter82} can be generalized in a straightforward manner to obtain
the eigenvalue expression and associated Bethe ansatz-type equations.

At $u=-(2s-1)\lambda$ the $R$-matrix $R^{(2s,2s)}(u)$ becomes proportional
to the exchange operator $P$, with proportionality constant $\prod_{j=1}^{2s}
\sinh(j \lambda)$. According to Eq.\ (\ref{eqn:ham}) the Hamiltonian for the
spin chain is then given by
\begin{equation}
\left.H=t^{(2s)}_{2s}(u)^{-1} \frac{d}{du} t^{(2s)}_{2s}(u)
\right|_{u=-(2s-1)\lambda} = \prod_{j=1}^{2s} \sinh(j \lambda)^{-1}
\left( \sum_{j=1}^{N-1} h_{j,j+1} + h^{F}_{N,1} \right),
\label{eqn:hamxxz}
\end{equation}
with two-site interaction $h_{j,j+1}=R_{ij}^{(2s,2s)}(-(2s-1)\lambda)P_{ij}$.
The Hamiltonian in terms of $U_q(su(2))$ generators is very complicated, but
using Eq.\ (\ref{eqn:fss}) the boundary conditions can be simply written as
\begin{equation}
S^z_{N+1} = -S^z_1, \hspace{20pt} S^{\pm}_{N+1} = S^{\mp}_1.
\end{equation}

\section{Transfer matrix diagonalization}
\setcounter{equation}{0}

In this section we will obtain the eigenvalue expression $\Lambda^{(2s)}_1(u)$
for the transfer matrix $t^{(2s)}_1(u)$, from which the
eigenvalue expressions for the transfer matrix $t^{(2s)}$ and the Hamiltonian
(\ref{eqn:hamxxz}) follow.  For this purpose we will require explicit
expressions for the non-zero matrix elements of $R^{(1,2s)}(u)$, which can be
obtained by interchanging $S^i$ and $\sigma^i$ in (\ref{eqn:rl1}):
\begin{eqnarray}
R^{(1,2s)}(u)_{1\;j}^{1\;j} & = & a_j(u), \nonumber\\
R^{(1,2s)}(u)_{2\;j}^{2\;j} & = & a_{n+1-j}(u), \nonumber\\
R^{(1,2s)}(u)_{1\;j+1}^{2\;j} & = & b_j(\lambda), \nonumber\\
R^{(1,2s)}(u)_{2\;j}^{1\;j+1} & = & b_j(\lambda),
\end{eqnarray}
where
\begin{eqnarray}
  a_j(u) & = & \sinh[u+(2s+1-j)\lambda],\nonumber\\
  b_j(\lambda) & = & \sqrt{\sinh(j\lambda)\sinh[\lambda(2s+1-j)]}.
\label{eqn:abdef}
\end{eqnarray}

\subsection{Pair-propagation through a vertex}

A key ingredient of Baxter's ``method of commuting transfer matrices''
\cite{Baxter82}
is the property of ``pair-propagation through a vertex''
for the $R$-matrix; ie.\ the existence of vectors $\hin$, $\hout$,
$\vin$ and $\vout$ such
that:\footnote{formulated here for $R$-matrices acting in $\IC^2 \otimes
\IC^{2s+1}$.}
\begin{equation}
 R(u)_{\mu \alpha}^{\nu \beta} \; \vin_{\beta} \hin_{\nu} =
 \vout_{\alpha} \hout_{\mu},
\label{eqn:ptv}
\end{equation}
for $\alpha,\beta \in \{1,\ldots,2s+1\}$ and $\mu,\nu \in \{1,2\}$.
Here $\vin_{\beta}$ are entries of the vector $\vin$ etc.

For the $R$-matrix $R^{(1,2s)}(u)$, the set of equations (\ref{eqn:ptv})
can be written as
\begin{eqnarray}
a_j(u) \; \vin_j \; \hin_1 + b_{j-1}(\lambda) \; \vin_{j-1} \;\hin_2 & = &
   \vout_j \; \hout_1,
\label{eqn:sys1}\\
a_{2s+2-j}(u) \; \vin_j \; \hin_2 + b_j(\lambda) \; \vin_{j+1} \; \hin_{1}
& = & \vout_j \; \hout_2,
\label{eqn:sys2}
\end{eqnarray}
where $j$ runs from $1$ to $2s+1$.
The first thing to note about the system of
equations (\ref{eqn:sys1}), (\ref{eqn:sys2})
is that it is linear in $\vin_j$ and $\vout_j$.
Therefore, consistency of (\ref{eqn:sys1}) and (\ref{eqn:sys2})
requires the vanishing of the
$(4s+2) \times (4s+2)$ determinant
\begin{equation}
\left|\begin{array}{cccc|cccc}a_1\hin_1 & & & &-\hout_1 & & & \\
                       b_1\hin_2 & \ddots & & & & \ddots & & \\
                       & \ddots & \ddots & & & & \ddots &\\
                       & & b_{2s}\hin_2 & a_{2s+1}
                           \hin_1& & & & -\hout_1\\ \hline
                       a_{2s+1}\hin_2 & b_1\hin_1 & & &-\hout_2 & & &\\
                       & \ddots & \ddots & & &  \ddots & &\\
                       & & \ddots & b_{2s}\hin_1 & & & \ddots &\\
                       & & & a_1\hin_2& & & & -\hout_2
             \end{array}\right|.
\label{eqn:det}
\end{equation}
An examination of small values of $s$ leads us to the following
conjecture:
\begin{claim}
Define $r^{{\rm in},{\rm out}}= h^{{\rm in},{\rm out}}_2/
  h^{{\rm in},{\rm out}}_1$. Then the determinant (\ref{eqn:det})
vanishes if
\begin{equation}
   \frac{r^{{\rm out}}}{r^{{\rm in}}}= \E^{2 \lambda \sigma},
\end{equation}
where $\sigma=-s,-s+1,\ldots,s$.
\end{claim}

We do not have a proof for general $s$, but this result has been important
in leading to Proposition 3 below, which is a stronger result and
will be proved. For that purpose,
we will require a few definitions: Let $q=e^{\lambda}$ and define the
$q$-numbers, factorials and binomial coefficients, respectively, by
\begin{eqnarray*}
[n] & = & \frac{q^n - q^{-n}}{q-q^{-1}} \;\; = \;\; \frac{\sinh(n\lambda)}
   {\sinh(\lambda)}\\
{[}n{]}! & = & [n][n-1]\cdots [1], \hspace{10pt} {\rm with}\; [0]!=1\\
\left[\begin{array}{c}m\\n\end{array} \right] & = & \frac{[m]!}{[m-n]![n]!},
\hspace{10pt} \left( m,n \in \IZ_{\geq 0},\; m\geq n \right).
\end{eqnarray*}
The $q$-binomial coefficients satisfy the following relation (a $q$-analogue
of the binomial theorem)
\begin{equation}
\sum_{k=0}^{2s+1} a^k \left[\begin{array}{c}2s+1\\k\end{array} \right]
=\prod_{j=-s}^s \left(1+a q^{2j}\right),
\label{eqn:qbt}
\end{equation}
which will be used later.
Introduce the functions $f^{(p)}(\sigma;u) =
\sum_{k=0}^p f^{(p)}_k(\sigma;u)$, with
\begin{equation}
f^{(p)}_k(\sigma;u) = (-e^{2 \lambda\sigma})^k
\left[\begin{array}{c}p\\k\end{array} \right]
\prod_{j=1}^k \sinh[u+(j-1)\lambda]\prod_{j=k+1}^p \sinh[u+(2s+j-p)\lambda].
\label{eqn:fdef}
\end{equation}
They have the useful properties summarised in the following:
\begin{lemma}
Let $a_j(u)$ and $b_j(\lambda)$ be defined as in (\ref{eqn:abdef}).
For any $\sigma\in \IC$, the
functions $f^{(p)}(\sigma;u)$ defined in (\ref{eqn:fdef}) satisfy
\begin{eqnarray}
f^{(p)}(\sigma;u) +
  e^{2\lambda\sigma}a_{2s+2-p}(u) f^{(p-1)}(\sigma;u)
   & = & a_1(u) f^{(p-1)}(\sigma;u-\lambda) ,\label{eqn:f1}\\
a_{p+1}(u) f^{(p)}(\sigma;u) + e^{2\lambda\sigma} b_p(u)^2
    f^{(p-1)}(\sigma;u)
   & = & f^{(p)}(\sigma;u-\lambda) a_1(u),\label{eqn:f2}
\end{eqnarray}
and
\begin{eqnarray}
\lefteqn{ a_1(u)a_{2s+1}(u) f^{(j-1)}(\sigma;u-\lambda)
   f^{(j-1)}(\sigma;u+\lambda)}\hspace{30pt}
\nonumber\\
& = & a_j(u) a_{2s+2-j}(u) f^{(j-1)}(\sigma;u)^2 -
   b_{j-1}(\lambda)^2 f^{(j)}(\sigma;u) f^{(j-2)}(\sigma;u).
\label{eqn:f4}
\end{eqnarray}
If $\sigma$ takes values only in the range $\{-s,-s+1,\ldots,s\}$, then
we also have
\begin{equation}
f^{(2s)}(\sigma;u-\lambda) = e^{2 \lambda\sigma} f^{(2s)}(\sigma;u).
\label{eqn:f3}
\end{equation}
\end{lemma}
\noindent
{ \bf Proof }
These relations can be proved using the definition of $f^{(p)}(\sigma;u)$. For
instance, Eq.\ (\ref{eqn:f1}) boils down to verifying
\begin{displaymath}
\left[\begin{array}{c}p-1\\k\end{array} \right] \sinh(u-\lambda) +
\left[\begin{array}{c}p-1\\k-1\end{array} \right] \sinh[u+(p-1)\lambda]=
\left[\begin{array}{c}p\\k\end{array} \right] \sinh[u+(k-1)\lambda].
\end{displaymath}
The proof of Eq.\ (\ref{eqn:f3}) is more interesting; the difference between
the two sides can be shown to be
\begin{displaymath}
\prod_{j=1}^{2s} \sinh[u+(j-1)\lambda] \sum_{k=0}^{2s+1}
  (-e^{2\lambda \sigma})^k \left[\begin{array}{c}2s+1\\k\end{array} \right].
\end{displaymath}
Upon application of the $q$-binomial theorem (\ref{eqn:qbt}), the sum above
simplifies to
\begin{displaymath}
\prod_{j=-s}^s \left(1-e^{2\lambda(\sigma+j)}\right)
\end{displaymath}
from
which the result immediately follows.

We are now ready to
state and prove the main result in this subsection:

\begin{proposition}
Let $\sigma \in \{-s,-s+1,\cdots,s\}$ and $f^{(p)}(\sigma;u)$
be defined as in (\ref{eqn:fdef}).  The relations
\begin{eqnarray}
\hin_2 & = & \frac{\hin_1 \hout_2}{\hout_1} \E^{2 \lambda \sigma},
\label{eqn:inp}\\
\vout_j & = & a_1(u) \;\frac{\hin_1 \vin_j}{\hout_1}
\; \frac{f^{(j-1)}(\sigma;u-\lambda)}{f^{(j-1)}(\sigma;u)},  \label{eqn:res1}\\
\vin_{j+1} & = & \frac{1}{b_j(\lambda)}\; \frac{\vin_j \; \hout_2}{\hout_1}
\; \frac{f^{(j)}(\sigma;u)}{f^{(j-1)}(\sigma;u)},
\label{eqn:res2}
\end{eqnarray}
with arbitrary $\hout_{1,2}$, $\hin_1$ and $v^{{\rm in,out}}_1$
solve the equations (\ref{eqn:sys1}) and (\ref{eqn:sys2}).
In other words, the $R$-matrix
$R^{(1,2s)}(u)$ {\em does} indeed have the property of
pair-propagation through a vertex.
\end{proposition}
\noindent
{\bf Proof }
The proof is by substitution of (\ref{eqn:inp}), (\ref{eqn:res1}) and
(\ref{eqn:res2}) directly into (\ref{eqn:sys1}) and (\ref{eqn:sys2}).
For ``generic'' values of $j$, these equations reduce to (\ref{eqn:f1})
and (\ref{eqn:f2}) which we have indicated above  are true.
For the ``end values'' $j=1$ and $j=2s+1$ equations
(\ref{eqn:sys1}) and (\ref{eqn:sys2}), respectively, degenerate by virtue
of the vanishing of $b_0(\lambda)$ and $b_{2s+1}(\lambda)$.
The former case can be
trivially seen to be true; the latter is true by virtue of (\ref{eqn:f3}).

For the special cases of $\sigma=\pm s$, the function $f^{(p)}(\sigma;u)$
simplifies to
\begin{equation}
  f^{(p)}(\pm s;u) = e^{\mp p u} \prod_{j=1}^{p}\sinh[(2s-j+1)\lambda],
\label{eqn:simp}
\end{equation}
with corresponding simplifications in the expressions for $\vin_j$ and
$\vout_j$ in Proposition 3. For the spin-$\smallfrac{1}{2}$ case,
we recover the six-vertex model results of \cite{Baxter82} by substituting
Eq.\ (\ref{eqn:simp}) into Proposition 3.

Let us indicate explicitly the $u$- and $\sigma$-dependence of the vectors
$\vin_{\sigma}(u)$ and
$\vout_{\sigma}(u)$. Equation (\ref{eqn:res2}) is a recursion relation for the
elements $\vin_{\sigma}(u)_j$, given the initial value $\vin_{\sigma}(u)_1$.
If we choose this initial value to be unity, then it is evident that
\begin{displaymath}
\vin_{\sigma}(u-\lambda)_j = \vin_{\sigma}(u)_j
    \frac{f^{(j-1)}(\sigma;u-\lambda)}
   {f^{(j-1)}(\sigma;u)}.
\end{displaymath}
It can then be shown from (\ref{eqn:res1})
that $\vout_{\sigma}(u)$ is simplify related to $\vin_{\sigma}(u)$ by
\begin{equation}
  \vout_{\sigma}(u) = a_1(u) \frac{\hin_1}{\hout_1} \vin_{\sigma}(u-\lambda).
\label{eqn:voutin}
\end{equation}

Proposition 3 can be interpreted in the following way: It
guarantees the existence of matrices
\begin{equation}
P = \left(\begin{array}{cc}
   \hout_1 & {*} \\
   \hout_2 & {*} \end{array} \right) \hspace{20pt}
P' = \left(\begin{array}{cc}
   \hin_1 & {*} \\
   \hin_2 & {*} \end{array} \right)
\label{eqn:ppp}
\end{equation}
such that the generalized similarity transformation
\begin{equation}
(P \otimes 1)^{-1} R_{12}^{(1,2s)}(u) (P' \otimes 1) = \left( \begin{array}{cc}
  \alpha(u) & \beta(u) \\
  \gamma(u) & \delta(u) \end{array} \right)
\label{eqn:sim}
\end{equation}
results in a matrix which is upper block-triangular when acting on a certain
vector in $\IC^{2s+1}$ to be specified.
To see this statement, pre-multiply (\ref{eqn:sim}) by $P \otimes 1$ on both
sides and act on $1 \otimes \vin_{\sigma}(u)$.
We end up with a $2\times 2$ matrix
equation with entries in $\IC^{2n+1}$. Component-wise, it reads
\begin{eqnarray}
\lefteqn{
\left(\begin{array}{cc}
R(u)_{1 j}^{1 k} \vin_{\sigma}(u)_k & R(u)_{1 j}^{2 k} \vin_{\sigma}(u)_k\\
R(u)_{2 j}^{1 k} \vin_{\sigma}(u)_k & R(u)_{2 j}^{2 k} \vin_{\sigma}(u)_k
   \end{array}\right)
\left(\begin{array}{cc}
\hin_1 & {*} \\ \hin_2 & {*} \end{array} \right) =}\hspace{40pt}\nonumber\\
& &
\left(\begin{array}{cc}
\hout_1 & {*} \\ \hout_2 & {*} \end{array} \right)
\left(\begin{array}{cc}
\alpha(u) \vin_{\sigma}(u)_j & \beta(u) \vin_{\sigma}(u)_j \\
\gamma(u) \vin_{\sigma}(u)_j & \delta(u) \vin_{\sigma}(u)_j \end{array}\right).
\label{eqn:expl}
\end{eqnarray}
A comparison with (\ref{eqn:ptv}) immediately indicates that
\begin{equation}
\gamma(u) \vin_{\sigma}(u)  =  0.
\label{eqn:cv}
\end{equation}
The action of $\alpha(u)$ and $\delta(u)$ on the vector $\vin_{\sigma}(u)$,
whose
components are determined from the recursion relation (\ref{eqn:res2}) with
$\vin_{\sigma}(u)_1=1$, is determined by examination of the explicit forms for
$\vin_{\sigma}(u)$ and $\vout_{\sigma}(u)$ given in Proposition 3:
With the help of (\ref{eqn:voutin}) we find that
\begin{equation}
\alpha(u) \vin_{\sigma}(u)  =  \frac{\hin_1}{\hout_1} \; a_1(u) \;
   \vin_{\sigma}(u-\lambda),
\end{equation}
whereas by taking determinants on both sides of (\ref{eqn:expl}), together
with the result (\ref{eqn:f4}), we find that
\begin{equation}
\delta(u) \vin_{\sigma}(u)  =  \frac{\hout_1 \det(P')}{\hin_1 \det(P)} \;
  a_{2s+1}(u) \;
   \vin(u+\lambda).
\label{eqn:dv}
\end{equation}

All the considerations in this subsection are local in nature. The results
are directly applicable to the transfer matrix with periodic boundaries,
generalizing the spin-$\smallfrac{1}{2}$ (six-vertex model) results of
\cite{Baxter82} and providing an alternative to the algebraic and analytic
Bethe ansatz \cite{Kirillov87}. In the next subsection we will specialize to
the boundary conditions specified by the twist matrices $F^{(l)}$ defined in
Section 3.

\subsection{Diagonalization of the auxiliary transfer matrix}

The main idea of the technique to diagonalize the transfer matrix
$t^{(2s)}_1(u)$ is to use Eqs. (\ref{eqn:sim}) and (\ref{eqn:cv}) --
(\ref{eqn:dv}) to construct functional equations satisfied by $t^{(2s)}_1(u)$.
We note that an attempt to do a generalized algebraic Bethe ansatz calculation,
along the lines of \cite{Takhtadzhan79} for the eight-vertex model, should
also probably use Eqs. (\ref{eqn:sim}) and (\ref{eqn:cv})-- (\ref{eqn:dv})
as a starting point.

Let us rewrite the auxiliary transfer matrix  in the form
\begin{equation}
t^{(2s)}_1(u)= \tr \left( P_{N+1}^{-1} \stackrel{a}{F^{(1)}}
P_{1} \left( P_{1}^{-1} R_{a1}^{(1,2s)}(u)
  P_2\right) \cdots \left( P_{N}^{-1} R_{aN}^{(1,2s)}(u)
  P_{N+1}\right)\right),
\end{equation}
where the matrices $P_k$ are understood to act non-trivially only in the
auxiliary space $V_a$. These matrices are chosen to be
\begin{eqnarray}
& & P_{N+1} = \left(\begin{array}{cc}
   1 & {*} \\
   r & {*} \end{array} \right), \hspace{15pt}
P_N = \left(\begin{array}{cc}
   1 & {*} \\
   r e^{-2\lambda \sigma_N} & {*} \end{array} \right),\hspace{15pt} \cdots
\nonumber\\
& & P_{2} = \left(\begin{array}{cc}
   1 & {*} \\
   r e^{-2\lambda \sum_{i=2}^{N}\sigma_i} & {*} \end{array} \right),
\hspace{15pt}
P_{1} = \left(\begin{array}{cc}
   r & {*} \\
   r^2 e^{-2\lambda \sum_{i=1}^{N}\sigma_i} & {*} \end{array} \right),
\end{eqnarray}
where $\sigma_i$, for $i=1,\ldots,N$ is any set of variables taking
values in  $\{-s,-s+1,\ldots, s\}$.  If, in addition, $r$ is chosen to satisfy
\begin{equation}
r^2 = \exp\left(2 \lambda \sum_{i=1}^{N} \sigma_i\right),
\label{eqn:rsig}
\end{equation}
then, by direct verification, we have the relation
\begin{equation}
P_{N+1}^{-1} \left(\begin{array}{cc}0 & 1\\1 & 0 \end{array}\right)
P_{1} = \left(\begin{array}{cc}1 & {*}\\ 0 & -\frac{\det(P_{1})}{\det(P_{N+1})}
\end{array}\right).
\end{equation}
Let us denote $\bigboldsigma=(\sigma_1,\ldots,\sigma_N)$.
{}From the results (\ref{eqn:ppp}) to (\ref{eqn:dv}) we deduce that
\begin{equation}
t^{(2s)}_1(u) \; y_{\boldsigma}(u) = \frac{1}{r} a_1(u)^N
y_{\boldsigma}(u-\lambda) - r a_{2s+1}(u)^N y_{\boldsigma}(u+\lambda),
\label{eqn:ty}
\end{equation}
where $y_{\boldsigma}(u)=v^{(1)}_{\sigma_1}(u)\otimes \cdots \otimes
v^{(N)}_{\sigma_N}(u)$ is a vector in
$\left(\IC^{2s+1}\right)^{\otimes N}$, with the components of
$v^{(k)}_{\sigma_k}(u)$ determined from the recursion relations
\begin{eqnarray}
v^{(k)}_{\sigma_k}(u)_1 & = & 1, \nonumber\\
v^{(k)}_{\sigma_k}(u)_{j+1} & = & v^{(k)}_{\sigma_k}(u)_j \; \frac{r
e^{-2\lambda( \sigma_1 + \cdots + \sigma_k)}}{b_j(\lambda)} \;
\frac{f^{(j)}(\sigma_k;u)}{f^{(j-1)}(\sigma_k;u)}.
\label{eqn:vrec}
\end{eqnarray}
Note that the $r$'s in Eq.\ (\ref{eqn:ty}) are $\bigboldsigma$-dependent
because of Eq.\ (\ref{eqn:rsig}); they can be removed by re-scaling
$y_{\boldsigma}(u)$ by a $u$-dependent factor.
The result can be summarised thus:
\begin{proposition}
Let $t^{(2s)}_1(u)$ be the transfer matrix defined by (\ref{eqn:tml}). The
vector
\begin{equation}
q_{\boldsigma}(u) = e^{\sum_{i=1}^N \sigma_i u} v^{(1)}_{\sigma_1}(u)\otimes
\cdots \otimes v^{(N)}_{\sigma_N}(u),
\end{equation}
where $v^{(k)}_{\sigma_k}(u)$ is determined by
(\ref{eqn:vrec}) with $r=\exp(\lambda \sum_{i=1}^N \sigma_i)$,
satisfies the equation
\begin{equation}
t^{(2s)}_1(u) \; q_{\boldsigma}(u) = a_1(u)^N q_{\boldsigma}(u-\lambda) -
a_{2s+1}(u)^N q_{\boldsigma}(u+\lambda).
\end{equation}
\end{proposition}

The vector $q_{\boldsigma}(u)$ in Proposition 4 is well-defined for any choice
of $\bigboldsigma=(\sigma_1,\dots,\sigma_N)$, with $\sigma_j \in
\{-s,-s+1,\ldots,s\}$.
Therefore the matrix $Q_R(u)$, whose columns are formed from the collection
of such vectors ($(2s+1)^N$ of them altogether), satisfies the relation
\begin{equation}
t^{(2s)}_1(u) \; Q_R(u) = a_1(u)^N Q_R(u-\lambda) -
a_{2s+1}(u)^N Q_R(u+\lambda).
\label{eqn:tqr}
\end{equation}
We will now require the following property of
$t^{(2s)}_1(u)$ which is due to crossing-symmetry of the
corresponding $R$-matrix:
\begin{lemma}
The transfer matrix $t^{(2s)}_1(u)$ satisfies the relation
\begin{equation}
   {}^t t^{(2s)}_1(-u-2\lambda s) = (-1)^{N-1} t^{(2s)}_1(u).
\label{eqn:cr}
\end{equation}
\end{lemma}
\noindent
{\bf Proof }
The $R$-matrix $R^{(1,2s)}(u)$ has the crossing relation
\begin{equation}
R^{(1,2s)}_{12}(u) = \stackrel{1}{V} {}^{t_2} R^{(1,2s)}_{12}(-u-2\lambda s)
\stackrel{1}{V},
\label{eqn:cross}
\end{equation}
with $V=\sigma^+ - \sigma^-$, which can be verified directly from Eq.
(\ref{eqn:rl1}). By the properties of the trace, we can write the lhs of
(\ref{eqn:cr}) as
\begin{eqnarray*}
 &  & \tr_a {}^{t_a t_1 \ldots t_N}
 \left( R^{(1,2s)}_{a1}(-u-2\lambda s) \cdots R^{(1,2s)}_{aN}(-u-2\lambda s)
  \stackrel{a}{F^{(1)}} \right)\\
 & = & \tr_a \left( \stackrel{a}{F^{(1)}}
  {}^{t_a t_N} R^{(1,2s)}_{aN}(-u-2\lambda s) \cdots
  {}^{t_a t_1} R^{(1,2s)}_{a1}(-u-2\lambda s) \right),
\end{eqnarray*}
which upon application of Eq. (\ref{eqn:cross}) becomes
\begin{eqnarray*}
  \cdots & =& \tr_a \left( \stackrel{a}{F^{(1)}}
  {}^{t_a}(\stackrel{a}{V^{-1}}  R^{(1,2s)}_{aN}(u) \stackrel{a}{V^{-1}})
  \cdots {}^{t_a}(\stackrel{a}{V^{-1}}
   R^{(1,2s)}_{a1}(u) \stackrel{a}{V^{-1}} ) \right) \\
  &=& \tr_a \left(\stackrel{a}{V^{-1}} R^{(1,2s)}_{a1}(u) \stackrel{a}{V^{-1}}
   \right) \cdots \left(\stackrel{a}{V^{-1}} R^{(1,2s)}_{aN}(u)
   \stackrel{a}{V^{-1}}\right) \stackrel{a}{F^{(1)}}.
\end{eqnarray*}
Upon using the properties $\left(V^{-1}\right)^2=-1$ and $V^{-1} F^{(1)}
V^{-1} = F^{(1)}$, the assertion follows.

With the relation (\ref{eqn:cr}), together with $a_1(-u-2\lambda s) =
- a_{2s+1}(u)$ and $a_{2s+1}(-u-2\lambda s) = -a_1(u)$ which follow from
the definition (\ref{eqn:abdef}), we conclude
from Eq. (\ref{eqn:tqr}) that
\begin{equation}
Q_L(u) \; t^{(2s)}_1(u)  = a_1(u)^N Q_L(u-\lambda) -
a_{2s+1}(u)^N Q_L(u+\lambda),
\label{eqn:tql}
\end{equation}
where we have defined $Q_L(u) = {}^t Q_R(-u-2\lambda s)$. The matrices
$Q_L(u)$ and $Q_R(u)$ obey the crucial ``commutation relations''
\begin{equation}
Q_L(u) Q_R(v) = Q_L(v) Q_R(u).
\label{eqn:comlr}
\end{equation}
This follows from the following result which we formulate as a conjecture:
\begin{conjecture}
For any choice  of $\sigma_i,\sigma'_i$ taking values in $\{-s,\ldots,s\}$,
the inner product
\begin{equation}
{}^t q_{\boldsigma'}(-u-2\lambda s) \cdot q_{\boldsigma}(v)
\label{eqn:symm}
\end{equation}
with the vector $q_{\boldsigma}(u)$ defined in Proposition 4, is symmetric in
$(u,v)$.
\end{conjecture}
Due to the tensor product structure of the vector $q_{\boldsigma}(u)$, the
inner product (\ref{eqn:symm}) assumes the form
\begin{equation}
{}^t q_{\boldsigma'}(-u-2\lambda s) \cdot
q_{\boldsigma}(v)=e^{(-u-2s\lambda)\sum_{i=1}^N
\sigma'_i + v \sum_{i=1}^N \sigma_i} \prod_{k=1}^N
G^{(k)}_{\sigma_k' \sigma_k}(u,v),
\label{eqn:gsimp}
\end{equation}
where
\begin{displaymath}
G^{(k)}_{\sigma_k' \sigma_k}(u,v) \equiv
 v^{(k)}_{\sigma_k'}(-u-2\lambda s) \cdot v^{(k)}_{\sigma_k} (v).
\end{displaymath}
However, from the recursion relations (\ref{eqn:vrec})
together with $r=e^{\lambda \sum_{i=1}^N \sigma_i}$, we have
\begin{equation}
v^{(k)}_{\sigma_k}(u)_{j+1} = e^{j \lambda \sum^k \boldsigma}
\frac{f^{(j)}(\sigma_k; u)}{b_1(\lambda) \cdots b_j(\lambda)},
\end{equation}
where $\sum^k \bigboldsigma$ denotes  $\sum_{i=k+1}^N \sigma_i -\sum_{i=1}^{k}
\sigma_i$, giving rise to
\begin{equation}
G^{(k)}_{\sigma_k' \sigma_k}(u,v) = \sum_{j=0}^{2s} \frac{e^{2j\lambda
\sum^k(\boldsigma+ \boldsigma')}}{\left(b_1(\lambda)\cdots
b_j(\lambda)\right)^2}
f^{(j)}(\sigma_k;v) f^{(j)}(\sigma_k';-u-2s\lambda).
\label{eqn:Gss}
\end{equation}
If we restrict $\sigma_i,\sigma'_i$ to take values only in the set $\{\pm s\}$
then $G^{(k)}_{\sigma_k' \sigma_k}(u,v)$ in Eq.\ (\ref{eqn:gsimp}) simplifies
and it can be proved (see Appendix A) that the inner product
 (\ref{eqn:gsimp}) is indeed
symmetric in $(u,v)$. For $s=\frac{1}{2}$ this is, of course, the whole story
\cite{Batchelor95}. For general $s$ we have not been able to construct a proof
of Conjecture 1 but
we have confirmed the result on a computer for a range of values of $s$ and
$N$.

We now require a technical assumption:
\begin{claim}
The matrix $Q_R(u)$ is non-singular for generic values of $u$. In particular,
it can be inverted at the point $u=u_0$.
\end{claim}
Define now the matrix $Q(u) = Q_R(u) Q_R^{-1}(u_0)$. With the property
(\ref{eqn:comlr}) and the above technical assumption (which is standard in the
present method \cite{Baxter82}, but is presumably difficult to prove),
we arrive at the key result that
\begin{eqnarray}
t^{(2s)}_1(u) \; Q(u) & = &  Q(u) \; t^{(2s)}_1(u) \nonumber\\
& = & a_1(u)^N Q(u-\lambda) -
a_{2s+1}(u)^N Q(u+\lambda),
\label{eqn:func}
\end{eqnarray}
together with $Q(u) Q(v) = Q(v) Q(u)$. In other words, we have constructed
a matrix $Q(u)$ which commutes with the transfer matrix
$t^{(2s)}_1(u)$, and with $Q(v)$ for $v$ different from $u$. This allows us
to choose the diagonal representation for all the matrices appearing in the
functional equation (\ref{eqn:func}), which then becomes a relation for
the eigenvalues of the transfer matrix.

\subsection{Bethe ansatz-type equations}

We can deduce the functional form of $Q(u)$ in its diagonal representation
(written also as $Q(u)$, with abuse of notation)
by examining Eq.\ (\ref{eqn:func}) in the limits $u \rightarrow \pm \infty$.
{}From Eq.\ (\ref{eqn:rl1}) the dominant term in the large $u$ limit in each
entry of $R^{(l,1)}(u)$ is
$\sim e^u$. Therefore from Eq.\ (\ref{eqn:tml}) the dominant term in each entry
of $t^{(2s)}_1(u)$ is $\sim e^{(N-1)u}$ (rather than $e^{Nu}$ because of the
twist matrix $F^{(l)}$). On general grounds the large $u$ behaviour of $Q(u)$
is argued to be $\sim e^{ru}$. Subsituting into Eq.\ (\ref{eqn:func}) we find
that $r$ must be $sN$. A similar argument holds for $u\rightarrow -\infty$
and we are led to the form
\begin{eqnarray*}
Q(u) & \sim & e^{-s N u} + \cdots + e^{s N u} \\
     & = & e^{-s N u} ( c_0 + c_1 e^u + \cdots + c_{2sN} e^{2sNu}),
\end{eqnarray*}
which can be parametrized alternatively as
\begin{equation}
Q(u) = \prod_{j=1}^{2sN} \sinh\left( \frac{u-u_j}{2}\right),
\end{equation}
after taking out an irrelevant overall $u$-independent factor.
The eigenvalue expression for the transfer matrix can then be
written as
\begin{eqnarray}
\Lambda^{(2s)}_1(u)& = & \sinh(u+2 s \lambda)^N \prod_{j=1}^{2sN} \frac{
   \sinh[\smallfrac{1}{2}(u-u_j-\lambda)]}
   {\sinh[\smallfrac{1}{2}(u-u_j)]}\nonumber\\
 &  - & \sinh(u)^N
  \prod_{j=1}^{2sN}  \frac{
\sinh[\smallfrac{1}{2}(u-u_j+\lambda)]}
  {\sinh[\smallfrac{1}{2}(u-u_j)]}.
\label{eqn:ev}
\end{eqnarray}
The requirement of analyticity of $\Lambda^{(2s)}_1(u)$ then imposes the
following conditions on the ``Bethe ansatz roots'' $u_j$:
\begin{equation}
\left( \frac{\sinh(u_k+2s\lambda)}{\sinh(u_k)}\right)^N = \prod_{j=1}^{2sN}
 \frac{\sinh[\smallfrac{1}{2}(u_k-u_j+\lambda)]}
      {\sinh[\smallfrac{1}{2}(u_k-u_j-\lambda)]}, \hspace{10pt} k=1,\ldots,2sN.
\label{eqn:ba}
\end{equation}
In the spin-$\smallfrac{1}{2}$ case, we recover from (\ref{eqn:ev}) and
(\ref{eqn:ba}) the eigenvalue expression and Bethe ansatz equations obtained in
\cite{Batchelor95}.
For general $s$, substitution of (\ref{eqn:ev}) into the generating function
(\ref{eqn:gf}) yields the eigenvalue expression for the transfer matrix
$t^{(2s)}_l(u)$ in terms of the Bethe ansatz roots $u_j$, namely
\begin{eqnarray}
\Lambda^{(2s)}_l(u) & = &
 A_l(u)^N \prod_{k=1}^{2sN} \frac{\sinh[\smallfrac{1}{2}(u-u_k-\lambda)]}
   {\sinh[\smallfrac{1}{2}(u-u_k+(l-1)\lambda)]} + (-)^{j+l}
   \sum_{j=1}^{l-1} A_j(u)^N \times \nonumber\\
& & \prod_{k=1}^{2sN}
  \frac{\sinh[\smallfrac{1}{2}(u-u_k+l\lambda)]\sinh[\smallfrac{1}{2}(
   u-u_k-\lambda)]}{\sinh[\smallfrac{1}{2}(u-u_k+j \lambda)]
   \sinh[\smallfrac{1}{2}(u-u_k+(j-1)\lambda)]} + \nonumber\\
&& (-)^l
  A_0(u)^N \prod_{k=1}^{2sN} \frac{\sinh[\smallfrac{1}{2}(u-u_k+l \lambda)]}
   {\sinh[\smallfrac{1}{2}(u-u_k)]},
\label{eqn:eig2s}
\end{eqnarray}
where
\begin{equation}
  A_j(u) = \prod_{p=j}^{l-1}\sinh(u+p \lambda)
     \prod_{p=0}^{j-1}\sinh[u+(2s+p)\lambda].
\end{equation}

The eigenvalue expression for the spin-$s$ XXZ Hamiltonian (\ref{eqn:hamxxz})
follows from $\Lambda^{(2s)}_{2s}(u)$
in Eq.\ (\ref{eqn:eig2s}): We find the following expression
\begin{equation}
\left. \frac{d}{du} \log \Lambda^{(2s)}_{2s}(u) \right|_{u=-(2s-1)\lambda} =
N \sum_{j=1}^{2s} \coth(j \lambda) + \sum_{k=1}^{2sN} \frac{\sinh(s\lambda)}
{\sinh[\frac{1}{2}(2s\lambda+u_k)]\sinh(\frac{1}{2}u_k)},
\label{eqn:evham}
\end{equation}
with the Bethe ansatz roots $u_k$ determined from Eq.\ (\ref{eqn:ba}).

We close this section with a few remarks on more general non-diagonal
twists for the transfer matrix. Namely, consider
\begin{equation}
\widehat{t}_1^{(2s)}(u) = \tr_a R^{(1,2s)}_{a1}(u) \cdots R^{(1,2s)}_{aN}(u)
\stackrel{a}{\widehat{F}^{(1)}},
\end{equation}
with
\begin{displaymath}
\widehat{F}^{(1)}
  \equiv \left(\begin{array}{cc}0 & \alpha\\ \beta & 0\end{array}
\right).
\end{displaymath}
We find the relationship
\begin{equation}
\det \left( \widehat{t}_1^{(2s)}(u) - \sqrt{\alpha\beta} \lambda
\right) \propto \det \left( t^{(2s)}_1(u) - \lambda \right)
\label{eqn:other}
\end{equation}
between the characteristic polynomials for the two transfer matrices,
giving the eigenvalue expression
\begin{equation}
\widehat{\Lambda}^{(2s)}_1(u) = \sqrt{\alpha\beta} \Lambda^{(2s)}_1(u),
\label{eqn:wide}
\end{equation}
with $\Lambda^{(2s)}_1(u)$ given by Eq.\ (\ref{eqn:ev}).
The twist matrix $\widehat{F}^{(1)}$ generates recursively the twist matrices
$\widehat{F}^{(l)}$, with the matrix elements of $\widehat{F}^{(l)}$ given
by $\widehat{F}^{(l)}_{i,j}= \delta_{i+j,l+2} \alpha^{l+1-i} \beta^{i-1}$,
through the decomposition
\begin{equation}
D_{12}^{(l)} \stackrel{1}{\widehat{F}^{(l)}}
    \stackrel{2}{\widehat{F}^{(1)}} D_{12}^{(l) \; -1}
= \left( \begin{array} {cc}  \widehat{F}^{(l+1)} & 0\\
  0 & -\alpha \beta \widehat{F}^{(l-1)} \end{array} \right),
\end{equation}
which replaces Eq.\ (\ref{eqn:DFD}) when $\alpha,\beta$ are different from 1.
The auxiliary transfer matrix $\widehat{t}^{(2s)}_1$ is then relevant to the
twist matrix $\widehat{F}^{(2s)}$ for the spin-$s$ XXZ chain corresponding
to boundary conditions
\begin{equation}
S^z_{N+1} = - S^z_1, \hspace{10pt} S^{+}_{N+1} = \frac{\beta}{\alpha}
S^-_1, \hspace{10pt} S^{-}_{N+1} = \frac{\alpha}{\beta} S^+_1,
\end{equation}
generalizing Eq.\ (\ref{eqn:bchalf}) to arbitrary spin-$s$. By Eq.\
(\ref{eqn:wide}) the Hamiltonians corresponding to these boundary conditions
share the eigenvalue expression Eq.\ (\ref{eqn:evham}) and Bethe ansatz
equations Eq.\ (\ref{eqn:ba}).

\section{Discussion}

In this paper we have studied integrable toroidal boundary conditions for
vertex models and related spin chains, concentrating on boundary conditions
specified by non-diagonal twists. In particular, we have applied the
``T-Q'' functional relation method \cite{Baxter82} to obtain the transfer
matrix eigenvalue expression and Bethe ansatz-type equations for the spin-$s$
XXZ Heisenberg chain, thereby generalizing a recent result \cite{Batchelor95}
for spin-$\smallfrac{1}{2}$.
A detailed analysis of these Bethe ansatz equations should confirm various
conjectures available \cite{Alcaraz88,Baranowski90} for the critical
properties of these models.

The primary reason why the abovementioned functional relation method is
applicable is because the $R$-matrix giving rise to the
transfer matrix $t^{(2s)}_1$ has the ``pair-propagation
through a vertex'' property.  Such considerations make sense in the
first place because of the two-dimensional nature of
the auxiliary space $V_a$ associated with this transfer
matrix. For more general $R$-matrices such as those
given in Refs.\ \cite{Jimbo86,Bazhanov87}, where the auxiliary space is no
longer two-dimensional, the appropriate generalization is not so clear.
Therefore the diagonalization of the transfer matrices for such models
in the case of non-diagonal twists remains an open problem.

\vspace{30pt}
\noindent
{\Large \bf Acknowledgements}
\vspace{10pt}

\noindent
We thank R. J. Baxter for a collaboration on the six-vertex model which led
to this work. We are also grateful to V. Rittenberg for helpful comments
and suggestions. This work is supported by the Australian Research Council.

\vspace{30pt}
\noindent
{\Large \bf Appendices}
\renewcommand{\thesection}{\Alph{section}}
\renewcommand{\theequation}{\Alph{section}.\arabic{equation}}
\setcounter{section}{0}
\section{Partial proof of Conjecture 1}
\setcounter{equation}{0}

\noindent
In this appendix we study the inner product
${}^t q_{\boldsigma'}(-u-2\lambda s) \cdot \;q_{\boldsigma}(v)$ of Conjecture
1,
for the case when $\sigma_i, \sigma_i'$ are allowed to take only the values
$\pm s$. In this case, the function $f^{(p)}(\sigma;u)$ simplifies according
to (\ref{eqn:simp}), accompanied by a simplification of
$G^{(k)}_{\sigma_k' \sigma_k}(u,v)$ in (\ref{eqn:Gss}) to
\begin{equation}
G^{(k)}_{\sigma_k' \sigma_k}(u,v)  =   \sum_{j=0}^{2s} \left[\begin{array}{c}
 2s\\j\end{array}\right]e^{j\lambda \sum^k (\boldsigma + \boldsigma')}\;
   e^{j(2 \sigma'_k \lambda-\frac{\sigma_k v}{s} + \frac{\sigma_k' u}{s})}.
\end{equation}
Here $\left[\begin{array}{c} 2s\\j\end{array}\right]$ is the $q$-binomial
coefficient defined in the paragraph preceding Lemma 1. By an application
of the $q$-binomial theorem (\ref{eqn:qbt}) we obtain
\begin{equation}
G^{(k)}_{\boldsigma' \boldsigma}(u,v)  =
\prod_{j=-s+1/2}^{s-1/2}\left(1+ e^{2j\lambda + \lambda
  \sum^k(\boldsigma+\boldsigma')}\;e^{2 \sigma_k' \lambda-\frac{\sigma_k v}{s}
+
  \frac{\sigma_k' u}{s}}\right),
\end{equation}
and therefore
\begin{eqnarray}
\lefteqn{
{}^t q_{\boldsigma'}(-u-2\lambda s) \cdot q_{\boldsigma}(v) =}\nonumber\\ & &
\prod_{k=1}^N \prod_{j=-s+1/2}^{s-1/2} \left( e^{\frac{\sigma_k v}{2s}-
   \frac{\sigma_k' u}{2s}-\lambda \sigma_k'} +
   e^{-\frac{\sigma_k v}{2s}+\frac{\sigma_k' u}{2s}+\lambda \sigma_k'+
   2j\lambda + \lambda
  \sum^k (\sigma_i + \sigma'_i) }\right).
\label{eqn:ssp}
\end{eqnarray}
Suppose now
there exists a pair $\sigma_p, \sigma'_p$ such that $\sigma_p+\sigma'_p
=0$. Then the terms in Eq.\ (\ref{eqn:ssp}) which involve $\sigma_p$ can be
seen to be symmetric in $(u,v)$, whereas the terms which do not involve
$\sigma_p$ can be seen (after some relabelling) to be expressible in the form
(\ref{eqn:ssp}) with $N \rightarrow N-1$. Hence we need only consider the
case where $\sigma_i=\sigma'_i$ for all $i$, which simplifies to
\begin{eqnarray}
\lefteqn{G_N(\sigma_1,\ldots,\sigma_N) \equiv
 {}^t q_{\boldsigma}(-u-2\lambda s) \cdot q_{\boldsigma}(v)  =
e^{-2\lambda s \sum_{k=1}^N \sigma_k}} \nonumber\\
& & \times
 \prod_{k=1}^N \prod_{j=-s+1/2}^{s-1/2}\left( e^{\frac{\sigma_k}{2s}(v-u)}
+e^{\frac{\sigma_k}{2s}(u-v) + 2j \lambda + 2\lambda\left[
\sum_{i=k+1}^N \sigma_i - \sum_{i=1}^{k-1} \sigma_i \right] }\right).
\label{eqn:ss}
\end{eqnarray}
If $\sigma_k+\sigma_{k+1}=0$ then from Eq.\ (\ref{eqn:ss}) we can write
$G_N(\sigma_1,\ldots,\sigma_k,-\sigma_k,\ldots,\sigma_N)=
G_{N-2}(\sigma_1,\ldots, \widehat{\sigma_k}, \widehat{\sigma_{k+1}},\ldots,
\sigma_N)$ multiplied by a function symmetric in $(u,v)$. This is true for
$1\leq k < N$. The only remaining case is therefore $\sigma_1 = \cdots =
\sigma_N$. But from Eq.\ (\ref{eqn:ss}) we again have $G_N(\sigma_1,\ldots,
\sigma_{N-1},\sigma_1)= G_{N-2}(\sigma_2,\ldots,\sigma_{N-1})$ up to a function
symmetric in $(u,v)$. We now check explicitly that $G_1(\sigma_1)$ and
$G_2(\sigma_1,\sigma_2)$ are symmetric functions in $(u,v)$ for all choices
of $(\sigma_1,\sigma_2)$. By induction on $N$, it follows that Eq.\
(\ref{eqn:ss}) is symmetric in $(u,v)$ for all choices of $\sigma_i$, thereby
proving that the inner product (\ref{eqn:symm}) is symmetric in $(u,v)$ for all
choices of $\sigma_i, \sigma'_i \in \{\pm s\}$, as claimed in the text.

\noindent
\section{Relation to the Hofstadter problem}
\setcounter{equation}{0}

\noindent
The Bethe ansatz equations (\ref{eqn:ba})
for the case when $N=1$ have already made a surprising appearance
\cite{Wiegmann94} in a quantum mechanical context.
This is in relation to the so-called Hofstadter problem of Bloch
electrons on a square lattice in a magnetic field. More specifically, in the
one-electron problem the Hamiltonian can be expressed in terms of the
generators of the group of magnetic translations, which in turn can be
related to $U_q(su(2))$-generators at $q=e^{\I \pi P/Q}$ (with $P,Q$ mutually
prime integers) in the $Q=(2s+1)$-dimensional representation. The eigenvalue
expression and associated Bethe ansatz equations were obtained in
\cite{Wiegmann94} using the functional Bethe ansatz. In this Appendix we
recover these equations from our general $N$ results. On the one hand, it
is a check on our results. On the other, it serves to further elucidate
the connection found in Ref.\ \cite{Wiegmann94}.

Define then the ``Hamiltonian'' ${\cal H}$ by
\begin{equation}
{\cal H} = 2\:\I\: \tr_a R^{(1,2s)}(u-s\lambda) \stackrel{a}{
  \left(\begin{array}{cc} 0 & 1\\1 & 0 \end{array}\right) }.
\label{eqn:hamhof}
\end{equation}
Comparison with Eq.\ (\ref{eqn:tml})
reveals immediately that ${\cal H}$ is simply
\begin{equation}
{\cal H}=
2\:\I\:t_1^{(2s)}(u-s \lambda),
\end{equation}
where the auxiliary transfer matrix on the rhs is for lattice size $N=1$.
On the other hand, it is clear from Eqs.\ (\ref{eqn:rl1}) and
(\ref{eqn:hamhof}) that ${\cal H}$ takes the form
\begin{equation}
{\cal H} = 2\:\I \sinh(\lambda) \left(S^+ + S^-\right)
\label{eqn:spsmhof}
\end{equation}
in terms of $U_q(su(2))$ generators. From the eigenvalue expression Eq.\
(\ref{eqn:ev})
for the auxiliary transfer matrix, we find the eigenvalue $E$ of ${\cal H}$
to be (after a shift in $u_k$)
\begin{eqnarray}
E & = & 2\:\I\sinh(u+ s\lambda) \prod_{j=1}^{2s}
  \frac{\sinh[\frac{1}{2}(u-u_j-\lambda)]}{\sinh[\frac{1}{2}(u-u_j)]}
  \nonumber\\
 & - & 2\:\I\sinh(u- s\lambda) \prod_{j=1}^{2s}
\frac{\sinh[\frac{1}{2}(u-u_j+\lambda)]}{\sinh[\frac{1}{2}(u-u_j)]},
\end{eqnarray}
together with the Bethe ansatz equations
\begin{equation}
\frac{\sinh(u_k+s \lambda)}{\sinh(u_k-s\lambda)} =
\prod_{j=1}^{2s} \frac{\sinh[\frac{1}{2}(u_k-u_j+\lambda)]}
   {\sinh[\frac{1}{2}(u_k-u_j-\lambda)]}.
\label{eqn:baeN1}
\end{equation}
Writing $q=e^{\lambda}$, the Bethe ansatz equations (\ref{eqn:baeN1}) can be
written in the alternative form
\begin{equation}
\frac{e^{2 u_k} - q^{-2s}}{e^{2u_k}q^{-2s} -1} =
\prod_{j=1}^{2s} \frac{e^{u_k} q - e^{u_j}}{e^{u_k} - e^{u_j} q},
\hspace{10pt} k=1,\ldots,2s.
\label{eqn:baeodd}
\end{equation}
It is clear from Eq.\ (\ref{eqn:spsmhof}) that ${\cal H}$ is independent
of $u$. Evaluating $E$ at $u\rightarrow \infty$ we obtain the expression
\begin{equation}
E = -\I (q-q^{-1}) \sum_{j=1}^{2s} e^{u_j}.
\label{eqn:eigodd}
\end{equation}

The Hamiltonian ${\cal H}$ has the following interpretation: There exists
a $q$-difference operator realization of the $(2s+1)$-dimensional
representation of $U_q(su(2))$ given by
\begin{eqnarray}
q^{\frac{1}{2}S^z} \Psi(z) & = & q^{-s} \Psi(qz) \\
q^{-\frac{1}{2}S^z} \Psi(z) & = & q^s \Psi(q^{-1}z)\\
S^+ \Psi(z) & = & z(q-q^{-1})^{-1} \left( q^{2s} \Psi(q^{-1}z) - q^{-2s}
    \Psi(qz) \right)\\
S^- \Psi(z) & = & - z^{-1}(q-q^{-1})^{-1} \left( \Psi(q^{-1}z)-\Psi(qz)
   \right),
\end{eqnarray}
with $\Psi(z)$ belonging to the space of polynomials of degree $2s$. In the
$q \rightarrow 1$ limit, this reduces to the familiar differential operator
realization of $su(2)$. In the $q$-difference operator realization,
the eigenvalue equation for the Hamiltonian ${\cal H}$ can be written as
\begin{equation}
\I (zq^{2s} - z^{-1}) \Psi(q^{-1}z) + \I
 (z^{-1}-z q^{-2s}) \Psi(qz)  = E \Psi(z).
\label{eqn:qdiff}
\end{equation}
Let $q=e^{\I \pi P/Q}$ with $Q=2s+1$ and consider first the case when
$P$ is odd. We have $q^{2s}=-q^{-1}$ and the difference equation
(\ref{eqn:qdiff}) becomes
\begin{equation}
- \I (z q^{-1}+ z^{-1}) \Psi(q^{-1}z) +
 \I (z^{-1}+ z q) \Psi(qz)  = E \Psi(z),
\label{eqn:qdiff2}
\end{equation}
while the Bethe ansatz equations determining the eigenvalue $E$ become
\begin{equation}
\frac{z_k^2 + q}{z_k^2 q +1} = -
\prod_{j=1}^{Q-1} \frac{z_k q - z_j}{z_k - z_j q}, \hspace{10pt}
 k=1,\ldots,Q-1,
\label{eqn:baez}
\end{equation}
with $z_j \equiv e^{u_j}$.  When $P$ is even we have $q^{2s}=q^{-1}$ and
we will again arrive at the difference equation (\ref{eqn:qdiff2})
if we replace ${\cal H}$ in Eq.\ (\ref{eqn:spsmhof}) by
$\tilde {\cal H} = 2 \I \sinh(\lambda) \left(S^- - S^+\right)$.
This corresponds, in Eq.\ (\ref{eqn:hamhof}),
to replacing the twist matrix $\left(\begin{array}
{cc}0 & 1\\ 1 & 0\end{array}\right)$ by $\left(\begin{array}
{cc}0 & -1\\ 1 & 0\end{array}\right)$, which is also an integrable twist. By
Eq.\ (\ref{eqn:other}), we find that the eigenvalue expression is
$\I$ times Eq.\ (\ref{eqn:eigodd}) with the same set of
Bethe ansatz equations (\ref{eqn:baeodd}).
The Bethe ansatz equations simplify, upon setting $e^{u_j}\equiv \I z_j$,
to the equations (\ref{eqn:baez}) without the negative sign in front of the
product on the rhs.
The odd and even $P$ Bethe ansatz equations can therefore be unified as
\begin{equation}
\frac{z_k^2 + q}{z_k^2 q +1} = q^Q
\prod_{j=1}^{Q-1} \frac{z_k q - z_j}{z_k - z_j q}, \hspace{10pt}
 k=1,\ldots,Q-1,
\label{eqn:baehof}
\end{equation}
with the eigenvalue expression becoming
\begin{equation}
E= -\I(q-q^{-1}) \sum_{j=1}^{Q-1} z_j.
\label{eqn:eighof}
\end{equation}

The difference equation (\ref{eqn:qdiff2}) is one form of  Harper's equation,
which appears in the Hofstadter problem with magnetic flux
$\Phi=2\pi \smallfrac{P}{Q}$ per plaquette. Apart from a sign
factor\footnote{We are in agreement
with Ref.\ \cite{Faddeev94} on the sign discrepancy.} in the Bethe ansatz
equations, the solution (\ref{eqn:eighof}) and (\ref{eqn:baehof}) for this
problem was first obtained in \cite{Wiegmann94} using the functional
Bethe ansatz.

\end{document}